\documentstyle[tighten,preprint,prl,aps]{revtex} \textheight24cm 

\begin{document}
\draft

\title{Constraints on Parity-Even Time Reversal Violation in the
Nucleon-Nucleon System and Its Connection to Charge Symmetry Breaking.}
\author{Markus Simonius}
\address{Institut f\"ur Teilchenphysik, Eidgen\"ossische Technische
Hochschule, CH-8093 Z\"urich, Switzerland}
\date{}
\maketitle

\begin{abstract}{Parity-even time reversal violation (TRV) in the
nucleon-nucleon interaction is reconsidered. The TRV
$\rho$-exchange interaction on which recent analyses of measurements
are based is necessarily also charge-symmetry breaking (CSB). 
Limits on its strength $\bar{g}_\rho$ relative to regular
$\rho$-exchange are extracted from recent CSB experiments in
neutron-proton scattering. The result $\bar{g}_\rho\le 6.7\times
10^{-3}$ (95\% CL) is considerably lower than limits inferred
from direct TRV tests in nuclear processes. 
Properties of $a_1$-exchange and limit imposed by the neutron EDM are
briefly discussed.}
\end{abstract}


\narrowtext

Investigations of time-reversal violation (TRV) orders of magnitude
above the weak interaction scale and thus necessarily of non-weak
origin continue to enjoy popularity in nuclear and
nucleon-nucleon physics 
\cite{Boehm95,Herczeg95,Haxton-Hoering,Haxton-Hoering-Musolf,%
Engel,Huffman,Vorov,Beyer}.

Since otherwise its presence would long have been detected in
experimental investigations of parity violation, TRV above the weak
interaction level must be parity-even. This imposes severe
restrictions on the possible ways a TRV interaction can be
constructed. Indeed, contrary to the case of a parity-odd TRV
interaction which naturally arises in the $\theta$-term of QCD, no
``natural'' implementation of P-even TRV is possible
\cite{Simonius-steamboat,Kambor,Herczeg95,Engel96}.

Also in an effective theory at the hadronic level, parity
conservation imposes restrictions which render most of the
experimental searches for P-even TRV rather elusive. For boson
exchange interactions in the nucleon-nucleon system these constraints
where analyzed in Ref.\ \cite{Simonius-TRV}. The main features, deduced
for on-shell amplitudes, are:

(i) P-even TRV is restricted to partial waves with total angular
momentum $J\ge 1$.

(ii) Total angular momentum zero exchange (one $\pi$ or $\sigma$ etc.)
can not contribute.

(iii) Natural parity exchange ($\rho$) must be charged
and necessarily contains the nucleon-nucleon isovector charge
exchange operator $2i(\tau^+_1\tau^-_2-\tau^-_1\tau^+_2)=
(\vec{\tau}_1\times\vec{\tau}_2)^z$ and thus cannot contribute in the
nn or pp system \cite{Khriplovich}.

Restrictions due to parity conservation are found also directly in
elastic scattering of nucleons on a spin zero nucleus: After
decomposition into angular momentum and parity eigenstates the
scattering matrix is diagonal and thus symmetric and automatically
time-reversal invariant. This shows also that a T-odd interaction
introduced at the nucleon-nucleon level is suppressed in nuclei since
it is ineffective in the interaction of single nucleons with the spin
zero core \cite{footnote1}. This ineffectiveness, emphasized already
in Ref. \cite{Simonius-Wyler} has recently been verified also for
processes in heavy nuclei governed by the statistical model
\cite{Haxton-Hoering,Haxton-Hoering-Musolf}, for which high
sensitivity has long been claimed, as well as in other calculations
\cite{Engel,Vorov,Beyer}.

Since charged $\rho$ is the lightest meson meeting the constraints (ii)
and (iii) above, it provides the longest range possible P-even TRV one boson
exchange interaction between nucleons. Following \cite{Haxton-Hoering}
most recent analyses of P-even TRV in nuclear systems are therefor
based on the $\rho$-exchange interaction proposed in Refs.
\cite{Simonius-TRV,Simonius-Wyler}.

However, according to (iii) a P-even TRV $\rho$-exchange nucleon-nucleon
interaction is necessarily CS-odd as well as T-odd, and thus invariant
under combined application of both these symmetry operations. It is
the main purpose of this paper to analyze the consequences of this
fact and assess the limits imposed on the strength of this interaction
by recent experimental measurements and theoretical analyzes of charge
symmetry breaking (CSB) in neutron-proton scattering.

\paragraph*{TRV $\rho$-exchange interaction.}
The starting point is the P-even TRV $\rho^\pm NN$-coupling (effective
Lagrangian) 
\begin{equation} \label{rNN-TRV}
 -i\bar{g}_\rho {g_\rho\kappa\over 2M}
\bar{\psi}\sigma_{\mu\nu}q^\nu\left(\vec{\tau}\times
\vec{\rho}^\mu\right)^z \psi
\end{equation}
where $M$ is the nucleon mass and $q$ the momentum of the (emitted)
$\rho$-meson. TRV is implemented by the isospin structure which
contains only charged $\rho$'s and and is odd under charge conjugation
$C$. With the corresponding regular coupling
\begin{equation} \label{rNN-reg}
g_\rho \bar{\psi}[\gamma_\mu +i{\kappa\over 2M}
\sigma_{\mu\nu}q^\nu]\left(\vec{\tau}\cdot\vec{\rho}^\mu\right) \psi
\end{equation}
one obtains the TRV Born amplitude or momentum space potential 
\begin{eqnarray} \label{Vrhoq}
\widetilde{V}^{TRV}_\rho(q) &=& \bar{g}_\rho {g^2_\rho\kappa\over 2M}
(\vec{\tau}_1\times\vec{\tau}_2)^z 
{1\over m^2_\rho+|\vec{q}|^2} \nonumber \\
&& \times i\Big((\vec{p}_f+\vec{p}_i) \times \vec{q}\Big) \cdot
(\vec{\sigma}_1-\vec{\sigma}_2)
\end{eqnarray}
to lowest order in the momenta where $\vec{p}_i$ and $\vec{p}_f$ denote
initial and final state relative momentum, respectively, and
$\vec{q}=\vec{p}_f-\vec{p}_i$.
Fourier transformation then yields the corresponding
TRV configuration space potential
\begin{eqnarray} \label{Vrhor}
V^{TRV}_\rho(r)&=&\bar{g}_\rho g^2_\rho\kappa
(\vec{\tau}_1\times\vec{\tau}_2)^z
\vec{l}\cdot(\vec{\sigma}_1-\vec{\sigma}_2) \nonumber \\
&&\times{m_\rho^2\over M^2}\ {e^{-m_\rho\,r} \over 4\pi\,r}
\left( {1\over m_\rho\,r} + {1\over m^2_\rho\,r^2}\right).
\end{eqnarray}
In these equations $g_\rho=2.79$ is the usual $\rho NN$ coupling
constant and $\bar{g}_\rho$ parametrizes the relative strength of the
TRV coupling. For definiteness of the definition of $\bar{g}_\rho$
$\kappa$ is identified in the TRV interaction with its vector
dominance value $\kappa=\mu_V=3.7$, the anomalous isovector nucleon
magnetic moment, though a larger value is preferred in the strong
nucleon-nucleon interaction.

It is emphasized that the structure of the interaction in Eqs.\
(\ref{Vrhoq},\ref{Vrhor}) is completely fixed by the fact that the
total exchanged system has natural parity $\pi=(-1)^J$ where J ($>$0)
is the total angular momentum of the exchanged system
\cite{Simonius-TRV}. In fact the same TRV NN interaction is obtained
if the TRV coupling (\ref{rNN-TRV}) is replaced by
$\bar{g}_\rho g_\rho\bar{\psi}\gamma_\mu\left(\vec{\tau}\times
\vec{\rho}^\mu\right)^z \psi$.
Moreover the same spin-isospin structure would be obtained for
two $\pi$ exchange since two pions are always in a natural parity
state and they also could not contribute if carrying total spin zero
which in the regular interaction gives their dominant contribution.

\paragraph*{Elastic neutron-proton scattering.}
The observables of interest are the single-spin observables $A^n$,
$A^p$, $P^n$, and $P^p$ where $A$ denotes the analyzing power for
neutrons (n) or protons (p) polarized perpendicularly to the
scattering plane and $P$ the corresponding polarization of outgoing
nucleons for unpolarized beam and target. In the
absence of TRV and CSB these observables are all equal:
$A^n=A^p=P^n=P^p$. T-odd effects are measured by
$\Delta^p \equiv P^p-A^p$ and
$\Delta^n \equiv P^n-A^n$
and CS-odd effects by
$\Delta A \equiv A^n-A^p$.
If time-reversal and charge-symmetry are broken by a purely T-odd {\em
and} CS-odd interaction as introduced above, 
i.~e. in the absence of any {\em other} (CS-even) T-odd or (T-even) CS-odd
effects, then
$P^n-A^p = P^p-A^n=0$
and
$\Delta A = \Delta_T^p = -\Delta_T^n \equiv \Delta$.

Of course CS is broken also otherwise, without violation of TRI, by
electromagnetic as well as quark mass effects which lead to a regular
CSB contribution $\Delta A^{\text{TRI}}$ to $\Delta A$ which is
supposed to be understood and calculable \cite{CSBtheor}. Thus the
actually measured quantity is
$ 
\Delta A = \Delta A^{\text{TRI}} + \Delta
$ 
where $\Delta$ is the contribution of the a TRV {\em and} CSB term of
interest in the present context.

Three experiments exist on CSB in neutron-proton scattering where
$\Delta A$ was measured \cite{477,183,347}. Their results, together
with theoretical predictions of $\Delta A^{\text{TRI}}$ \cite{CSBtheor},
based on updated experimental and theoretical values collected in Ref.
\cite{347} are listed in Table \ref{EXP}.

\paragraph*{Analysis of limits on $\bar{g}_\rho$.}
In order to deduce the limits on $\bar{g}_\rho$ listed in the last
columns of Table \ref{EXP} helicity amplitudes
$\langle\gamma\delta|T|\alpha\beta\rangle$
\cite{Bystricky,Simonius-LNP} are used. All (P-even) T-odd {\em and}
CS-odd effects are contained in one P-even combination $N$ of
single-spinflip helicity amplitudes defined by
\begin{equation} \label{N} 
\begin{array}{r@{\ }l} 
N =1/8\ \big(\ \langle + +|T|+ -\rangle -\langle - -|T|- +\rangle &\\
{}+ \langle + +|T|- +\rangle -\langle - -|T|+ -\rangle &\\
{}+ \langle + -|T|+ +\rangle -\langle - +|T|- -\rangle &\\
{}+ \langle - +|T|+ +\rangle -\langle + -|T|- -\rangle &\big).
\end{array}
\end{equation}
Its symmetry properties should be compared to those of the
corresponding regular spin-flip amplitude $M_5$ \cite{Bystricky} (the
relative minus sign between the two columns is imposed by parity
conservation in both cases while the second and third line have the
opposite sign in the regular amplitude $M_5$). The symmetry properties
of all other helicity amplitudes ($M_1$\dots$M_4$) \cite{Bystricky} are
completely fixed by parity conservation. By convention the first
helicity refers to the neutron and the second to the proton on both
sides of the matrix elements. Taking this into account and adjusting
the normalization, the $\rho$-exchange TRV Born amplitude (\ref{Vrhoq})
translates into 
\begin{equation} \label{Nborn}
N^{\text{Born}}=i\,\bar{g}_\rho {g^2_\rho\kappa\over 4\pi M}\
{2p^2\sin\theta \over m^2_\rho+2p^2(1+cos\theta)}
\end{equation}
where $\theta$ is the center-of-mass scattering angle.

Using the definition (\ref{N}) the evaluation of
the TRV and CSB asymmetry $\Delta$ yields
$ 
\Delta = 4\,\text{Im} [N^*C]/\sigma
$ 
where $k$ is the center of mass momentum, $\sigma=d\sigma/d\Omega$
stands for the normalizing differential cross section, and
\begin{equation}
C={1\over2}(-M_1+M_2+M_3+M_4) = {1\over 2k}(-H_1+H_2)
\end{equation}
in terms of the regular helicity amplitudes $M_i$ or dimensionless
``Virginia'' amplitudes $H_i$ \cite{Bystricky,Arndt}.

In order to proceed with the analysis consider the partial wave
decomposition of $N$ which contains only
singlet-triplet $l=J$ transitions with $J\ne 0$ \cite{Simonius-TRV}
\begin{equation} \label{NJ}
N(\theta)={1\over 2k}\sum_{J\ge 1}(2J+1)N_J d^J_{10}(\theta)
\end{equation}
where $d^J_{10}(\theta)$ is the appropriate Wigner rotation function
and $N_J$ are TRV and CSB $l=J$ singlet-triplet partial wave
transition amplitudes whose normalization will play no role. To
lowest order in the TRV interaction and in the absence of inelasticity
they are real by unitarity \cite{Simonius-TRV} up to a phase factor
$i\exp\{i(\delta^s_J+\delta^t_J)\}$ where $\delta^s_J$ and $\delta^t_J$
are the (experimentally known) singlet and triplet $l=J$ scattering
phases. Neglecting small inelasticities ($\sim 1\%$ at 347 and $\sim
10\%$ at 477 MeV) the dependence of $\Delta$ 
on $\bar{g}_\rho$ due to the TRV $\rho$-exchange interaction
(\ref{Vrhoq}) can therefor be written
\begin{equation}
\Delta=\bar{g}_\rho \sum_{J \ge 1} \alpha_J K_J(\theta)
\equiv\bar{g}_\rho \bar{K}(\alpha)
\end{equation}
where $\bar{K}(\alpha)=\sum_{J \ge 1} \alpha_J K_J(\theta)$
and
\begin{equation} \label{alpha}
\alpha_J = 
\text{Re}[N_J e^{-i(\delta^s_J+\delta^t_J)}/N_J^{\text{Born}}]
= \left|N_J/N_J^{\text{Born}}\right|
\end{equation}
is the ratio between the exact first order amplitude $N_J$ calculated
from the TRV $\rho$-exchange interaction (e.~g. in DWBA) and the
corresponding amplitude $N_J^{\text{Born}}$ obtained in Born
approximation. All the kinematics, known strong amplitudes and phases
are contained in
\begin{eqnarray} \label{KJ}
K_J(\theta)={g^2_\rho\kappa\over \pi M \sigma}\,
{(-1)^{J-1}(2J+1)\over \sqrt{J(J+1}}\sqrt{w^2-1}\,Q_J^1(w)
\nonumber \\ 
\times d^J_{10}(\theta)\,
\text{Re}\left[e^{-i(\delta^s_J+\delta^t_J)}C(\theta)\right]
\end{eqnarray}
where $w={m_\rho^2\over 2 k^2}+1$ and $Q_J^1(w)$ is the usual
associated Legendre function of the second kind arising from the
partial wave projection of the Born amplitude (\ref{Nborn}).
$\alpha_J\le 1$ accounts for the reduction in the matrix elements due
to strong short range repulsion between the nucleons (which is not
expected to change the sign).

Numerical values of the coefficients $K_J$ for angles and energies of
available CBS experiments \cite{477,183,347}, evaluated with phase
parameters and amplitudes obtained from \cite{SAID}, are listed in Table
\ref{THEOR} together with the corresponding values for
$\bar{K}(\alpha)=\Delta/\bar{g}_\rho$. The 183 MeV results are
averaged over the angular range of the measurement as the experimental
value $\Delta A$ \cite{theta}. In order to exhibit the sensitivity to
short range correlations between neutron and proton, $\alpha_1$ is
varied between 1 and 1/3 while the others are kept fixed at
$\alpha_J=1$. Note the disproportionate sensitivity of
$\Delta/\bar{g}_\rho$ on $\alpha_1$ due to destructive contribution of
the higher partial waves. Since the lowest partial waves that
contribute are $p$-waves, $\alpha_1={1\over 2}$ is considered to be
adequate rather than ${1 \over 3}$ which would be adequate for
$s$-waves in similar situations. The framed entries in Table
\ref{THEOR} are therefor used for the determination of limits on
$\bar{g}_\rho$.

The final limit
\begin{equation} \label{limit}
|\bar{g}_\rho| \le 6.7\times 10^{-3},\ \ 95\%\text{ CL}
\end{equation}
($|\bar{g}_\rho| \le 4.4\times 10^{-3}$, 80\% CL) is obtained from the
weighted average of the three values for $\bar{g}_\rho$ extracted from
the three CSB measurements in Table \ref{EXP}. This limit is
considerably lower than those obtained so far
from the analysis of nuclear processes
\cite{Haxton-Hoering,Haxton-Hoering-Musolf,Engel,Huffman,Vorov}.
It is also slightly better than the constraints obtained from atomic
electric dipole moments (EDM's)
\cite{Haxton-Hoering,Haxton-Hoering-Musolf}. 
Only the limit $\sim 10^{-3}$ due to the neutron EDM obtained in
Ref.\ \cite{Haxton-Hoering-Musolf} is lower numerically though its
dependence on the elusive parity violating $\pi NN$ coupling constant
$f_\pi$ \cite{HH} renders it somewhat uncertain
\cite{Haxton-Hoering-Musolf}. 

How reliable are the theoretical calculations on which this limit is
based? The only appreciable uncertainty of the calculation of $\Delta$
as a function of $\bar{g}_\rho$ presented here lies in the somewhat
crude estimate of $\alpha_J$ which could (and should) be improved by
explicit calculations of the matrix elements. More subtle are the
theoretical predictions of $\Delta A^{\text{TRI}}$ which have to be
subtracted from the experimental values for $\Delta A$ in order to
deduce experimental limits on $\Delta$. In order to assess their
reliability several aspects should be noted: The neutron being
uncharged, the main contributions are from the electromagnetic
interaction between proton and the magnetic moment of the neutron, the
n-p mass difference in the usual Boson exchange interaction, and
$\rho-\omega$ mixing. Only the latter has some inherent uncertainty
stemming from the extrapolation of the mixing parameter off the mass
shell. Fortunately, however, this affects only the analysis of the
183 MeV measurement \cite{183}. The others are, at the angles
measured, essentially insensitive to this contribution and thus to its
uncertainty \cite{347,Miller-Vanoers}. Since the limit (\ref{limit})
is dominated by the 347 MeV result it is therefor not affected by this
uncertainty. An uncertainty of 10\% say in $\Delta A^{TRI}$ leads to
an overall systematic error of 0.0020 in the average $\bar{g}_\rho$
extracted. Added in quadrature this increases the error of
$\bar{g}_\rho$ by a factor 1.2 and the final upper limits to 0.0077
(95\% CL) and 0.0051 (80\% CL). Careful scrutiny of the calculations
of $\Delta A^{TRI}$ in view of the present analysis would clearly be
desirable.

Finally it is reiterated that the structure of the interaction
analyzed is unique up to radial or $|q|$-dependence
\cite{Simonius-TRV}. Any total natural parity ($\pi = (-1)^J$)
exchange contribution must be of the same form.
Since total spin zero exchange cannot contribute, the only way to
circumvent the limit given is to turn to unnatural parity exchange,
generically $a_1(1260)$ or (at least) three-$\pi$ exchange
\cite{Simonius-TRV}.

In nuclear tests a TRV $a_1$-exchange interaction would be even more
elusive than $\rho$-exchange due to its even shorter range (higher
mass) and its smaller regular coupling. Moreover, in contrast to
$\rho$-exchange, there is ample freedom in isospin structure
\cite{Simonius-TRV} in order to escape limits.
On the other hand, the P-even TRV coupling of the $a_1$ to the
neutron is of the EDM form \cite{Simonius-Wyler,Sudarshan}
\begin{equation} \label{a1nn}
{f_T \over 2M} \bar{\psi}\sigma_{\mu\nu}\gamma^5 q^\nu a_1^\mu\psi.
\end{equation}
This allows one to to connect it directly to the neutron EDM $d_n$
using vector meson dominance of the electromagnetic current (leading
to the identification $\kappa=\mu_V=3.7$)
in conjunction with (parity violating) $\rho-a_1$ mixing.
Neglecting a similar $\omega$-contribution this yields
\begin{equation}
d_n \approx f_T{e \over 2M} {h_{\rho A}\over g_\rho }
= 3.8\times 10^{-15} f_T h_{\rho A}\ e\cdot cm 
\end{equation}
where $h_{\rho A}\approx 10^{-6}$ represents weak $\rho-a_1$ mixing
which gives rise to the so-called factorization contribution to parity
violating $\rho NN$ coupling \cite{DDH} whose magnitude is
experimentally verified \cite{Simonius-Tedaldi,HH}. With the
experimental limit $d_n\le 1.1\times 10^{-25}\ e\cdot cm$ 95\% CL
\cite{DN}, this leads to the limit $f_T\le 3\times 10^{-5}$ in the TRV
$a_1nn$ coupling (\ref{a1nn}).

A roughly ten times lower limit ($4\pi\beta_{qq} < 3.8 \times 10^{-6}$)
is presented in Ref.~\cite{Conti} for axial vector exchange at the quark
level based on loop calculation of the neutron EDM with
non-renormalizable coupling of the form (\ref{a1nn}). Their conclusion
that the P-even TRV nucleon-nucleon interaction cannot exceed
$10^{-4}$ of the usual weak interaction, however, is based on the
additional assumption that the exchanged object must have a mass $\mu
> 100$ GeV (which does not affect the EDM) and thus rests on the
explicit assumption that the interaction must involve the weak scale.
The object of the present analysis, on the other hand, is to analyse
evidence obtained from experiments and long range physics directly
without imposing such additional restrictions.

Note finally that limits obtained for axial vector exchange should not
be taken over to vector exchange (and vice versa) without further
examination due to their different symmetry properties
\cite{Simonius-TRV}. Unless both are isovector
interactions they cannot even generate one another without
intervention of additional isospin violation.

I would like to thank W.T.H. van Oers for bringing the insensitivity of
the TRIUMF CSB measurements \cite{477,347} to $\rho-\omega$ mixing to
my attention, and P.\ Herczeg for valuable comments.

\begin{table}
\widetext
\caption
{Summary of CSB experiments in elastic np scattering, theoretical
calculations of regular contributions $\Delta A^{\text{TRI}}$, and
ensuing limits for the TRV contribution $\Delta$. At 183 MeV the
asymmetries are averaged over the angular range indicated, for the
others they are at the zero-crossing angle of the (average) analyzing
power. The last three columns give the values of $\bar{g}_\rho$
obtained from $\Delta$ using the framed sensitivities
$\bar{K}(\alpha_1={1\over 2})$ given in Table \protect\ref{THEOR} and
corresponding 80\% and 95\% confidence limits.}
\label{EXP}
\begin{tabular}{ccr@{$\,\pm\,$}lcr@{$\,\pm\,$}lr@{$\,\pm\,$}ldd}
$T_{\text{lab}}[\text{MeV}]$ & $\theta_{\text{cm}}[^o]$ &
	\multicolumn{2}{c}{$\ \Delta A[10^{-4}]$} &
	$\Delta A^{\text{TRI}}[10^{-4}]$ &
	\multicolumn{2}{c}{$\ \Delta[10^{-4}]$} &
	\multicolumn{2}{r}{$\bar{g}_\rho[10^{-4}]$} &
	$|\bar{g}_\rho|$ 80\% CL &
	$|\bar{g}_\rho|$ 95\% CL \\ \hline
183 & 82.2$-$116.1 & 34.8 & 7.4 & 33 & 1.8 & 7.4 &
	20 & 83 & $\le$ 0.011 & $\le$ 0.017 \\
347 & 72.8       & 59   & 10  & 53 & 6   & 10 &
	22 & 37 & $\le$ 0.0056 & $\le$ 0.0084 \\
477 & 69.7       & 47   & 23  & 55 & $-$8  & 23 &
	$-$27 & 77 & $\le$ 0.011 & $\le$ 0.016 \\
\multicolumn{7}{l}{Weighted average of $\bar{g}_\rho$ and
corresponding limits:} & 
	14 & 31 & $\le$ 0.0044 & $\le$ 0.0067
\end{tabular}
\end{table}

\begin{table}
\widetext
\caption
{Coefficients $K_J$ for the partial wave contributions to $\Delta$ and
sensitivities $\bar{K}= \Delta/\bar{g}_\rho = \sum_{J \ge 1} \alpha_J
K_J$ of $\Delta$ on $\bar{g}_\rho$ calculated as indicated with
reduction factors $\alpha_1=1$, ${1 \over 2}$, and ${1 \over 3}$ for
the lowest contributing partial wave $J=1$, and $\alpha_J=1$ for
$J>1$.  The case with $\alpha_1=1$ is the strong phase modified Born
approximation result. In the last column the unmodified Born
approximation result is given for comparison, demonstrating the
importance of including the right phases. The framed entries
($\alpha_1={1\over 2}$) are used for the extraction of limits on
$\bar{g}_\rho$ in Table \protect\ref{EXP}.}
\label{THEOR}
\begin{tabular}{ccrrrrcccc}
$T_{\text{lab}}[\text{MeV}]$ & $\theta_{\text{cm}}[^o]$ & $K_1\quad$ &
	$K_2\quad$ & $K_3\quad$ & $K_4\quad$ & $\bar{K}(\alpha_1=1)$ &
	$\bar{K}(\alpha_1={1 \over 2})$ &
	$\bar{K}(\alpha_1={1 \over 3})$ & $\bar{K}^{\text{Born}}$ \\ \hline
183 & 82.2$-$116.1 & 0.1679 & 0.0082 & $-$0.0027 & $-$0.0001 &
	0.173 & \fbox{0.089} & 0.061 & 0.317 \\
347 & 72.8 & 0.7921 & $-$0.1132 & $-$0.0205 & 0.0065 &
	0.665 & \fbox{0.269} & 0.137 & 1.195 \\
477 & 69.7 & 0.9833 & $-$0.1776 & $-$0.0303 & 0.0154 &
	0.789 & \fbox{0.297} & 0.133 & 1.440 
\end{tabular}
\end{table}


\begin{references}
\bibitem{footnote1}
In nuclear calculations one has to be careful not to loose the
destructive interference leading to this suppression in inadequate
approximations.
\bibitem{theta}
At 347 and 477 MeV the sensitivities at the angles of the
measurements are within about 10\% of their maximum. At 183 MeV,
however, one would gain almost a factor two around $75^0$; over the
angular range of the measurement $\bar{K}(\alpha_1={1\over 2})$
decreases from 0.156 to 0.011.
\end{references}
\end{document}